\documentclass{article}%

\usepackage{amsmath}
\usepackage{amsfonts}
\usepackage{amssymb}
\usepackage{graphicx}%
\newcommand{\bpartial}{\mathop{\partial\kern -4pt\raisebox{.8pt}{$|$}}}
\newcommand{\bra}{\mathopen{[\kern-1.6pt[}}
\newcommand{\ket}{\mathclose{]\kern-1.5pt]}}
\newcommand{\bbra}{\mathopen{[\kern-2.2pt[\kern-2.3pt[}}
\newcommand{\bket}{\mathclose{]\kern-2.1pt]\kern-2.3pt]}}

\makeindex
\begin{document}

\title {\large{ \bf Energy-momentum tensors for non-commutative Abelian Proca field}}

\vspace{3mm}

\author {  \small{ \bf F. Darabi }\hspace{-1mm}{
\footnote{Corresponding author, e-mail:f.darabi@azaruniv.edu
}} \\
{\small{\em Department of Physics, Azarbaijan Shahid Madani University}}
{\small{\em   53714-161, Tabriz, Iran  }}\\
{\small{\em Research Institute for Astronomy and Astrophysics of
Maragha (RIAAM)}}
{\small{\em  55134-441 Maragha, Iran   }}
 \\and \\
\small{ \bf F. Naderi}\hspace{-1mm}{ \footnote{ e-mail:f.naderi@azaruniv.edu }}\\ { \small {\em Young Researchers and Elite Clube, Marand Branch, Islamic Azad University, Marand , Iran}}}

\maketitle
\begin{abstract}
We study two different possibilities of constructing the energy-momentum tensors for non-commutative Abelian Proca field, by using (i) general Noether theorem and (ii) coupling to a weak external gravitational field. Both energy-momentum tensors are not traceless due to the violation of Lorentz invariance in non-commutative spaces. In particular, we show that the obtained energy density of the latter case coincides exactly with that of obtained by Dirac quantization method. 
\end{abstract}

\section{Introduction}
The idea of non-commutative geometry was put forward by Alain Connes \cite{connes}, and has received much attention in recent years in field theory \cite{SW, ncft} following the development of the superstring theory. In such theories, the parameter of non-commutativity, $\theta_{\mu\nu}$, which is assumed to be a constant antisymmetric tensor, characterizes the non-commutative geometry through the commutation relation $[\hat{x}_{\mu},\hat{x}_{\nu}]=i\theta_{\mu\nu}$, and leads to a violation of Lorentz invariance \cite {LD1,LD}. Particularly, it is expected that the corresponding energy-momentum tensor needs not to be traceless as a consequence of violating the scale invariance. The energy-momentum tensor in non-commutative field theories has been studied recently from different points of view \cite{NC maxwell, EMNC, EMNC2}.

The formulation of non-commutative Maxwell filed based on Seiberg-Witten map \cite{SW} has already been
done in Refs.\cite{NC maxwell}, \cite{Kruglov}. With this motivation, we have presented the Dirac quantization of Abelian Proca field in  non-commutative space in \cite{proca prev}. The aim of present work is to construct the possible energy-momentum tensors in non-commutative space for Abelian Proca field theory, in order to work out the aspects of energy-momentum tensor and investigate the consistency with the Hamiltonian constraint system formalism given in \cite{proca prev}.

The organization of the paper is as follows. In section 2, we demonstrate the generalized non-commutative Lagrangian of Abelian Proca model at cubic order in electromagnetic field and obtain its filed equations. Then, the
corresponding Hamiltonian constraint system is represented. In section 3, we first study the construction of energy-momentum tensor via general Noether theorem giving a tensor which is neither symmetric nor traceless. Afterwards, we construct the energy-momentum tensor by coupling the Proca model to a weak gravitational field whose energy-momentum tensor is symmetric but not traceless due to the violation of Lorentz and dilation symmetry in non-commutative spaces. The paper ends with a brief conclusion.
\section{Hamiltonian constraint system approach }

The non-commutative Lagrangian for Abelian Proca field is given by \cite{proca prev}
 \begin{eqnarray}\label{1}
 \hat{\cal L}=-\frac{1}{4}F_{\mu\nu}F_{\mu\nu}+\frac{1}{8}\theta_{\alpha\beta}F_{\alpha\beta}F^{2}_{\mu\nu}-\frac{1}{2}\theta_{\alpha\beta}F_{\mu\alpha}F_{\nu\beta}F_{\mu\nu}+\frac{1}{2} m^{2}(A^{2}_{\mu}-\theta_{\alpha\beta}A_{\alpha}(\partial_{\beta}A_{\mu}+F_{\beta\mu})A_{\mu}),
 \end{eqnarray}
 where $A_{\mu}\equiv(\vec{A},iA_{0})$ is the electromagnetic potential and $F_{\mu\nu}=\partial_{\mu}A_{\nu}-\partial_{\nu} A_{\mu}$ is the field strength tensor. The Lagrangian can be rewritten as
 \begin{eqnarray}\label{2}
 \hat{{\cal L}}&=&\frac{1}{2}(E^2-B^2)(1+\vec{\theta} \cdot \vec{B})-(\vec{\theta} \cdot \vec{E})(\vec{E}
 \cdot \vec{B})+\frac{m^2}{2}(-A_0^2+A_i^2)
 \\ \nonumber
 &+&\frac{m^2}{4}(\vec{\theta} \times \vec{A})\cdot \vec{\nabla}(A_0^2)-\frac{m^2}{2}[(\vec{\theta} \times \vec{A})\cdot \vec{E}]A_0+3\frac{m^2}{4}[(\vec{\theta} \cdot \vec{B})A_j^2-(\vec{\theta} \cdot \vec{A})(\vec{A} \cdot \vec{B})],
 \end{eqnarray} 
 where the magnetic induction field is $B_{i}=\frac{1}{2}\epsilon_{ijk}F_{jk}$ and the electric filed is $E_{i}=iF_{i4}$.
 The Euler-Lagrange equations give the corresponding equations of motion as follows
 \begin{eqnarray}\label{3}
 \partial_{\rho}F_{\rho \sigma}&-&\frac{1}{4}\partial_{\rho}(\theta_{\rho \sigma}F_{\mu
 \nu}^2)-\frac{1}{2}\partial_{\rho}(\theta_{\alpha \beta}F_{\alpha
 \beta}F_{\rho \sigma})+\partial_{\rho}(\theta_{\sigma \beta}F_{\nu
 \beta}F_{\rho \nu})\\ \nonumber
 &-&\partial_{\rho}(\theta_{\rho \beta}F_{\nu
 \beta}F_{\sigma \nu})+\frac{1}{2}\partial_{\rho}(\theta_{\alpha \beta}F_{\rho
 \alpha}F_{\sigma \beta})-\frac{1}{2}\partial_{\rho}(\theta_{\alpha \beta}F_{\sigma
 \alpha}F_{\rho \beta})\\ \nonumber
 &-&\frac{m^2}{2}\theta_{\sigma \beta}(\partial
 _{\beta}{A}_{\mu}+F_{\beta \mu})A_{\mu}-\frac{m^2}{2}\theta_{\alpha \beta}(\partial
 _{\beta}{A}_{\sigma}+F_{\beta \sigma})A_{\alpha}\\ \nonumber
 &+&m^2\partial_{\rho}(\theta_{\alpha \rho}A_{\alpha}A_{\sigma})-\frac{m^2}{2}\partial_{\rho}(\theta_{\alpha \sigma}A_{\alpha})A_{\rho}+m^2A_{\sigma}=0.
 \end{eqnarray}
Following Dirac formalism for quantizing the Hamiltonian constraint systems \cite{Dirac}, we construct the Hamiltonian formalism of non-commutative Abelian proca filed. We first evaluate the conjugate momenta as follows
 \begin{eqnarray}\label{4}
 \pi_{i}=\frac{\partial\hat{\cal L}}{\partial(\partial_{0}A_{i})}=-E_{i}(1+\theta\cdot B)+(\theta\cdot E)B_{i}+(E\cdot B)\theta_{i}+\frac{m^{2}}{2}(\theta\times A)_{i}A_{0},
 \end{eqnarray}
 \begin{eqnarray}\label{5}
 \pi_{0}=\frac{\partial\hat{\cal L} }{\partial(\partial_{0}A_{0})}=0.
 \end{eqnarray}
 Equation \eqref{5} is, by definition, a primary constraint, $\varphi_{1}\equiv\pi_{0}\approx0$, where $\approx$ stands for weak equation \cite{Dirac, H}.
The Legendre transformation, ${\cal H}_0=\pi_{\mu}\partial_0 A_{\mu}-{\cal L}$, gives the canonical Hamiltonian density 
\begin{eqnarray}\label{6}
\hat{{\cal H}}_0&=&\frac{1}{2}(E^2+B^2)(1+\vec{\theta} \cdot \vec{B})-(\vec{\theta} \cdot \vec{E})(\vec{E}
\cdot \vec{B})+\frac{m^2}{2}A_0^2-\frac{m^2}{2}A_i^2(1+\frac{3}{2}\vec{\theta} \cdot \vec{B})
\\ \nonumber
&-&\frac{m^2}{2}(\vec{\theta} \times \vec{A})_i (\partial_i A_0)A_0+3\frac{m^2}{4}(\vec{\theta} \cdot \vec{A})(\vec{A} \cdot \vec{B})-\pi_i \partial_i A_0,
\end{eqnarray}
 where we have used equation \eqref{4} to describe $E_{i}$ in terms of momentum $\pi_{i}$. Now, the consistency condition of the primary constraint \cite{Dirac}, namely $\dot{\varphi_{1}}=\{\varphi_{1},H_{0}\}=0$, results in the secondary constraint as
  \begin{eqnarray}\label{7}
\varphi_{2}=\partial_{i}\pi_{i}+m^{2}A_{0}+\frac{m^{2}}{2}\nabla\cdot(\theta\times A)A_{0}+\frac{m^{2}}{2}(\theta\times A)\cdot E \approx 0.
 \end{eqnarray}
Now, including the primary constraint with an arbitrary function $u(x)$ in the original Hamiltonian density (\ref{6}) gives the total Hamiltonian as \begin{eqnarray}\label{8}
H_{T}=H_{0}+\int u(x)\varphi_{1}d^{3}x.
 \end{eqnarray}
The consistency condition for the secondary constraint (\ref{7}) fixes the arbitrary function $u(x)$ \cite{proca prev} and no new secondary constraint arises.
Based on Dirac classification, the two constraints $\varphi_{1}$ and $\varphi_{2}$ are second class constraints because of their non vanishing Poison bracket. Defining the matrix of Poisson bracket of constraints
 \begin{eqnarray}\label{9}
C_{ij}=\{\varphi_{i}(x),\varphi_{j}(y)\}\neq0,
 \end{eqnarray}
 where the inverse matrix $C^{-1}_{ij}$ exist, the Dirac bracket is introduced by \cite{Dirac}
 \begin{eqnarray}\label{10}
  \{A(x,t),B(y,t)\}_{DB}=\{A(x,t),B(y,t)\}-\int\{A(x,t),\varphi_{i}(z,t)\}C^{-1}_{ij}(z,w)\{\varphi_{j}(w,t),B(y,t)\}d^{3}zd^{3}w.
  \end{eqnarray}
Then, the non zero Dirac brackets are obtained
  \begin{eqnarray}\label{11}
 \{\pi_{i}(x,t),A_{j}(y,t)\}_{DB}= -\delta_{ij}\delta(x-y),
 \end{eqnarray}
    \begin{eqnarray}\label{12}
 \{A_{0}(x,t),A_{j}(y,t)\}_{DB}=m^{-2}(1-\frac{1}{2}\nabla\cdot(\theta\times A))\partial_{j}(x)\delta(x-y)-\frac{1}{2}(\theta\times A)_{j}\delta(x-y),
 \end{eqnarray}
    \begin{eqnarray}\label{13}
 \{\pi_{i}(x,t),A_{0}(y,t)\}_{DB}=\frac{1}{2}[A_{0}\epsilon_{ijl}\theta_{l}\partial_{j}(y)-(\pi\times\theta)_{i}(y)]\delta(x-y).
 \end{eqnarray}
 According to Dirac's prescription for second class constraints, Dirac bracket of any operator with the constraints vanishes \cite{Dirac}. Therefore, all second class constraints can be set strongly to zero, and this will eliminate all unphysical degrees of freedom from the theory. Consequently, the physical Hamiltonian density in which the unphysical quantities are absent, is given by
  \begin{eqnarray}\label{14}
 \hat{\cal H}_{ph}&\equiv&\varepsilon\nonumber\\
 &=&\frac{1}{2}(\pi^2+B^2)+\frac{1}{2}(B^2-\pi^2)(\vec{\theta} \cdot \vec{B})+(\vec{\pi} \cdot \vec{\theta})(\vec{B} \cdot \vec{\pi})
 -\frac{m^2}{2}A_0^2
 \\ \nonumber
 &-&\frac{m^2}{2}A_i^2(1+\frac{3}{2}\vec{\theta} \cdot \vec{B})
 +\frac{m^2}{4}\vec{\nabla} \cdot (\vec{\theta} \times \vec{A})A_0^2+
 3\frac{m^2}{4}(\vec{\theta} \cdot \vec{A})(\vec{A} \cdot \vec{B}).
 \end{eqnarray}

\section{Energy-momentum tensor}
In this section, we will study the construction of energy-momentum tensors for non-commutative Abelian Proca field in classical level. Our starting point is the Lagrangian \eqref{1}
which is obtained upon using the Seiberg-Witten map \cite{SW}. General procedure of Noether theorem leads to a non-symmetric canonical conservative energy- momentum tensor as \cite{Landau}
\begin{eqnarray}\label{15}
T_{\mu\nu}^{can}=(\partial_{\nu}A_{\alpha})\left(\frac{\partial\cal L}{\partial(\partial_{\mu}A_{\alpha})}\right)-\delta_{\mu\nu}\cal L.
 \end{eqnarray}
To get gauge invariant energy-momentum tensor, gauge transformation is performed on it in the following form
 \begin{eqnarray}\label{16}
T_{\mu\nu}=T_{\mu\nu}^{can}+\Lambda_{\mu\nu},
 \end{eqnarray}
 where $\partial_{\mu}\Lambda_{\mu\nu}=0$, and
  \begin{eqnarray}\label{17}
\Lambda_{\mu\nu}=(\partial_{\alpha}A_{\nu})\left(\frac{\partial\cal L}{\partial(\partial_{\mu}A_{\alpha})}\right).
 \end{eqnarray}
Using this expression, we may obtain conservative gauge invariant energy- momentum tensor for source free Proca field in non-commutative space as 
\begin{eqnarray}\label{18}
T_{\mu\nu}&=&-F_{\mu\alpha}F_{\nu\alpha}(1-\frac{1}{2}\theta_{\gamma\beta}F_{\gamma\beta})+\frac{1}{4}\theta_{\mu\alpha}F_{\nu\alpha}F^{2}_{\rho\beta}-\theta_{\mu\beta}F_{\gamma\nu}F_{\rho\beta}F_{\gamma\rho}-(F_{\mu\alpha}F_{\nu\gamma}+F_{\nu\alpha}F_{\mu\gamma})\theta_{\alpha\beta}F_{\gamma\beta}\nonumber\\
&-&\frac{m^{2}}{2}(2\theta_{\gamma\mu}F_{\nu\alpha}A_{\alpha}A_{\gamma}-\theta_{\gamma\alpha}F_{\nu\alpha}A_{\gamma}A_{\mu})-\delta_{\mu\nu}\cal L.
\end{eqnarray}
Although the tensor is symmetric in the limit $\theta\rightarrow 0$, it is still non-symmetric in first order of $\theta$.
The components of $T_{\mu\nu}$ are given as follows
    \begin{eqnarray}\label{19}
T_{44}&=&\frac{1}{2}(E^{2}+B^{2})(1+\theta \cdot B)-(E\cdot\theta)(\theta\cdot E)-\frac{m^{2}}{2}A_{i}^{2}(1+\frac{3m^{2}}{2}(\theta\cdot B))\nonumber\\
&+&\frac{3m^{2}}{4}(\theta\cdot A)(A\cdot B)+\frac{m^{2}}{2}A_{0}^{2}-\frac{m^{2}}{2}(\theta\times A)(\partial_{i}A_{0})A_{0},
 \end{eqnarray}
 \begin{eqnarray}\label{20}
 T_{i4}&=&-i([1+\theta\cdot B](E\times B)_{i}+\frac{1}{2}(B^{2}-E^{2})(E\times \theta)_{i}\nonumber\\
 &+&m^{2}[-A_{i}^{2}(E\times \theta)_{i}+(E\times\theta)_{i}(\theta\cdot A)-\frac{1}{2}E.(\theta\times A)A_{i}]),
 \end{eqnarray}
\begin{eqnarray}\label{21}
T_{4i}=-i([1+\theta\cdot B](E\times\theta)_{i}+(E\cdot B)(B\times\theta)_{i}-\frac{m^{2}}{2}A_{0}[(\theta\cdot B)A_{i}-(A\cdot B)\theta_{i}]),
 \end{eqnarray}
 \begin{eqnarray}\label{22}
T_{ij}&=&E_{i}E_{j}+B_{i}B_{j}-\delta_{ij}B^{2}+(\theta\cdot B)(2E_{i}E_{j}+B_{i}B_{j})+\frac{1}{2}(E^{2}+B^{2})B_{i}\theta_{j}\nonumber\\
&-&\frac{1}{2}\delta_{ij}(E^{2}+3B^{2})(\theta\cdot B)-(\theta\cdot E)E_{i}B_{j}-(E.B)(\theta_{i}E_{j}+\theta_{j}E_{i})-(E\times\theta)_{i}(B\times E)_{j}\nonumber\\
&-&m^{2}[(\theta\times A)_{i}E_{j}A_{0}+(\theta\times A)_{i}(A\times B)_{j}-\frac{1}{2}A_{i}A_{j}(\theta\cdot B)-\frac{1}{2}\theta_{i}A_{j}(A\cdot B)]-\delta_{i j}\cal L,
 \end{eqnarray}
and the trace of energy-momentum tensor is obtained as
 \begin{eqnarray}\label{23}
T_{\mu\mu}=(\theta\cdot B)(E^{2}-B^{2})-2(\theta\cdot E)(E\cdot B)+m^{2}(A_{0}^2+A_{i}^{2})-\frac{m^{2}}{2}E\cdot(\theta\times A)A_{0}-m^{2}(\theta\times A)_{i}(\partial_{i}A_{0})A_{0}.
 \end{eqnarray}
 Clearly, the energy-momentum tensor is not traceless even if $\theta\rightarrow0$.
 On the other hand, when $\cal L$ does not depend on $\partial_{\alpha}g^{\mu\nu}$, symmetric energy-momentum tensor can be generally derived from following procedure \cite{EMNC, Landau}
 \begin{equation}\label{24}
 T_{\mu\nu} = \left.\frac{2}{\sqrt{-g}}\,\frac{\delta (\sqrt{-g}~\cal \hat{L})}{\delta
   g^{\mu\nu}}\right|_{g^{\mu\nu}=\eta^{\mu\nu}},
 \end{equation}
where the action is given by the coupling of the non-commutative Proca filed to a weak external gravitational field as
\begin{eqnarray}\label{25}
\hat{\cal L}&=&-\frac{1}{4}F_{\mu\nu}F_{\rho\sigma}g^{\mu\rho}g^{\nu\sigma}(1-\frac{1}{2}\theta_{\alpha\beta}F_{\gamma\delta}g^{\alpha\gamma}g^{\beta\delta})
-\frac{1}{2}\theta_{\alpha\beta}F_{\mu\gamma}F_{\nu\gamma}F_{\rho\sigma}g^{\alpha\gamma}g^{\beta\delta}g^{\rho\mu}g^{\sigma\nu}\nonumber\\
&+&\frac{m^{2}}{2}(A_{\mu}A_{\nu}-\theta_{\alpha\beta}A_{\gamma}(\partial_{\sigma}A_{\nu}+F_{\sigma\nu})A_{\mu}g^{\gamma\alpha}g^{\sigma\beta})g^{\mu\nu},
\end{eqnarray}
Here $g$ denotes the determinant of the metric which will be set to the Euclidean metric at the end. Varying the Lagrangian \eqref{25} with respect to the metric tensor leads to the symmetric energy-momentum tensor as follows
   \begin{eqnarray}\label{26}
T^{sym}_{\alpha\beta}&=&-F_{\mu\alpha}F_{\nu\alpha}(1-\frac{1}{2}\theta_{\gamma\beta} F_{\gamma\beta})+\frac{1}{4}(\theta_{\mu\alpha}F_{\nu\alpha}+\theta_{\nu\alpha}F_{\mu\alpha})F_{\rho\beta}^{2}-\theta_{\mu\beta}F_{\gamma\nu}F_{\rho\beta}F_{\gamma\rho}\nonumber\\
&-&\theta_{\nu\beta}F_{\gamma\mu}F_{\rho\beta}F_{\gamma\rho}-(F_{\mu\alpha}F_{\nu\gamma}+F_{\nu\alpha}F_{\mu\gamma})\theta_{\alpha\beta}F_{\gamma\beta}+m^{2}A_{\mu}A_{\nu}\nonumber\\
&-&\frac{m^{2}}{2}[(\theta_{\mu\beta}A_{\nu}+\theta_{\nu\beta}A_{\mu})(\partial_{\beta}A_{\rho}+F_{\beta\rho})A_{\rho}+\theta_{\alpha\mu}A_{\alpha}(\partial_{\nu}A_{\rho}
+F_{\nu\rho})A_{\rho}\nonumber\\
&+&{\theta_{\alpha\nu}A_{\alpha}(\partial_{\mu}A_{\rho}+F_{\mu\rho})A_{\rho}+\theta_{\alpha\beta}A_{\alpha}(\partial _{\beta} A_{\nu}+F_{\beta\nu})A_{\mu}+\theta_{\alpha\beta}A_{\alpha}(\partial_{\beta}A_{\mu}+F_{\beta\mu})A_{\nu}]-\delta_{\mu\nu}\cal L},~~~
 \end{eqnarray}
where its components are given by
 \begin{eqnarray}\label{27}
T^{sym}_{i4}&=&-i[((1+\theta\cdot B)(E\times B)_{i}+\frac{1}{2}(B^{2}-E^{2})(E\times\theta)_{i}-m^{2}(\frac{1}{4}A_{0}^{2}+A_{j}^{2})(E\times\theta)_{i}\nonumber\\
&+&m^{2}(\theta\cdot A)(E\times A)_{i})-m^{2}A_{i}(A_{0}+\frac{1}{2}(\theta\times A)\cdot E-\frac{1}{2}(\theta\times A)_{i}\partial_{i}A_{0})\nonumber\\
&+&\frac{3}{4}(E\times\theta)_{i}A_{0}^{2}+\frac{3}{2}(A_{j}^{2}(E\times\theta)_{i}-(E\times A)_{i}(\theta\cdot A))+\frac{1}{2}A_{i}(\theta\times A)\cdot E\nonumber\\
&-&\frac{1}{2}\varepsilon_{ijk}\theta_{k}A_{0}^{2}\partial_{j}A_{0}+\frac{1}{2}(\theta\times A)_{i}(\partial_{0}A_{j})A_{j},
 \end{eqnarray}
  \begin{eqnarray}\label{28}
T^{sym}_{ij}&=&E_{i}E_{j}+B_{i}B_{j}-\frac{1}{2}\delta_{ij}B^{2}+(\theta\cdot B)(3E_{i}E_{j}+B_{i}B_{j})+\frac{1}{2}(E^{2}+B^{2})[B_{i}\theta_{j}+\theta_{i}B_{j}]\nonumber\\
&-&\frac{1}{2}(E^{2}+4B^{2})(\theta\cdot B)-(E\cdot B)[E_{i}\theta_{j}+\theta_{i}E_{j}]-(\theta\cdot E)(E_{i}B_{j}+E_{j}B_{i})\nonumber\\
&-&(E\times\theta)_{i}(B\times E)_{j}-(E\times\theta)_{j}(B\times E)_{i}\nonumber\\
&+&m^{2}[A_{i}A_{j}(1+3(\theta\cdot B))-\frac{3}{4}(A_{i}B_{j}+B_{i}A_{j})(\theta\cdot A)-\frac{1}{2}(A_{i}(E\times\theta)_{j}+A_{j}(E\times\theta)_{i})A_{0}\nonumber\\
&+&\frac{1}{4}(\varepsilon_{ilk}\theta_{k}A_{j}+\varepsilon_{jlk}\theta_{k}A_{i})\partial_{l}A_{0}^{2}-\frac{3}{4}[(\theta\times A)_{i}(A\times B)_{j}+(\theta\times A)_{j}(A\times B)_{i}\nonumber\\
&+&E_{i}(\theta\times A)_{j}+E_{j}(\theta\times A)_{i}]+\frac{3}{8}[(\theta\times A)_{i}\partial_{j}A_{0}^{2}+(\theta\times A)_{j}\partial_{i}A_{0}^{2}]-\frac{3}{4}(\theta_{i}A_{j}+\theta_{j}A_{i})(A\cdot B)]-\delta_{ij}\cal L,
 \end{eqnarray}
which represents the flow of the $i$th component of momentum in the $j$th direction (stress), and 
   \begin{eqnarray}\label{29}
T^{sym}_{44}&=&\frac{1}{2}(E^{2}+B^{2})(1+\theta\cdot B)-(\theta\cdot E)(E\cdot B)\nonumber\\
&+&\frac{m^{2}}{2}(-A_{0}^{2}+(\theta\times A)_{i}(\partial_{i} A_{0})A_{0}-A_{i}^{2}(1+\frac{3}{2}(\theta\cdot B))+\frac{3}{2}(\theta\cdot A)(A\cdot B)-(\theta\times A)\cdot E A_{0}).
 \end{eqnarray}
Substitution of \eqref{4} into (\ref{29}) yields
   \begin{eqnarray}\label{30}
T^{sym}_{44}&=&\frac{1}{2}(\pi^{2}+B^{2})+\frac{1}{2}(B^{2}-\pi^{2})(\theta\cdot B)-(\pi\cdot\theta)(B \cdot\pi)-\frac{m^{2}}{2}A_{0}^{2}+\frac{m^{2}}{2}(\theta\times A)_{i}(\partial_{i}A_{0})A_{0}\nonumber\\
&-&\frac{m^{2}}{2}[A_{i}^{2}(1+\frac{3}{2}(\theta\cdot B))]+\frac{3m^{2}}{4}(\theta\cdot A)(A\cdot B).
 \end{eqnarray}
This agrees with the energy density $\varepsilon$ (physical Hamiltonian density) which we have obtained by Dirac quantization of non-commutative Abelian Proca field in \eqref{14}.
Furthermore, the non zero trace of energy-momentum tensor is given by
 \begin{eqnarray}\label{31}
T^{sym}_{\mu\mu}=2(\theta\cdot B)(E^{2}-B^{2})-4(\theta\cdot E)(E\cdot B)+3m^{2}[A_{i}^{2}(\theta\cdot B)-(\theta\cdot A)(A\cdot B)+3(\theta\times A)_{i}(\partial_{i}A_{0}-E_{i})].
 \end{eqnarray}
 Obviously, there is a trace anomaly in first order of $\theta$.
 
Eventually, one observes that both calculated versions of the energy-momentum tensor are not traceless. In general, traceless energy-momentum tensor reflects the dilation invariance of theory, required in conformal field theory \cite{traceless}. However, the appeared trace anomaly in non-commutative space is due to the
violation of Lorentz and dilation symmetry \cite{LD}.

In particular, we are interested in the energy-momentum tensor of equation \eqref{26}, which beside admitting the physical energy density of Dirac quantization of Proca field \cite{proca prev}, agrees with energy-momentum tensor in commutative space \cite{proca commutative} in $\theta \rightarrow 0$ limit.
It is worth mentioning that if the Proca action could be regarded as a matter action in non-commutative general gravity, the trace anomaly would  then contribute to the cosmological constant as in Ref. \cite{d}.
\section{Conclusion}
We have studied the non-commutative Abelian Proca field, which is equivalent to massive Maxwells theory, by Dirac quantization formalism which includes second class constraints. Then, we have constructed possible energy-momentum tensors for Abelian Proca field in classical level. We have shown that the presence of non-commutativity parameter $\theta$ results in violation of Lorentz and dilation symmetry and therefore corresponding energy-momentum tensors are not traceless. We have also shown that the time-time component of $T^{sym}_{\mu\nu}$ which indicates the energy density agrees with the physical energy density obtained by Dirac quantization method.
\section*{Acknowledgment}
This work has been supported financially by Research Institute
for Astronomy and Astrophysics of Maragha (RIAAM) under research project NO.1/3252-43.
%

\end{document}